\documentclass[preprint,12pt]{elsarticle}

\usepackage[T1]{fontenc}
\usepackage[latin9]{inputenc}
\usepackage[english]{babel}
\usepackage{amsmath}
\usepackage{graphicx}
\usepackage[normalem]{ulem}
\usepackage{xcolor}
\usepackage{hyperref}
\usepackage{notoccite}

\journal{Physics Letters A}

\begin{document}

\begin{frontmatter}



\title{Light-field imaging from position-momentum correlations}


\author[uniba,infn]{Davide Giannella}
\author[uniba,infn]{Gianlorenzo Massaro}
\author[upol]{Bohumil Stoklasa} 
\author[uniba,infn]{\\ Milena D'Angelo\corref{cor}} 
\author[uniba,infn]{Francesco V. Pepe}

\address[uniba]{Dipartimento Interuniversitario di Fisica, Universit\`a degli studi di Bari, I-70126 Bari, Italy}
\address[infn]{INFN, Sezione di Bari, I-70126 Bari, Italy}
\address[upol]{Department of Optics, Palack\'y University, 77146 Olomouc, Czech Republic}

\cortext[cor]{milena.dangelo@uniba.it}

\begin{abstract}
Correlation plenoptic imaging (CPI) is a light-field imaging technique employing intensity correlation measurements to simultaneously detect the spatial distribution and the propagation direction of light. Compared to standard methods, in which light-field images are directly encoded in intensity, CPI provides a significant enhancement of the volumetric reconstruction performance in terms of both achievable depth of field and 3D resolution. In this article, we present a novel CPI configuration where light-field information is encoded in correlations between position and momentum measurements, namely, points on a given object plane and points of the Fourier plane of the imaging lens. Besides the fundamental interest in retrieving the properties of position-momentum correlation, the proposed scheme overcomes practical limitations of previously proposed setups, providing higher axial homogeneity and robustness with respect to the identification of reference planes. 
\end{abstract}







\end{frontmatter}

\section{Introduction}

In the rapidly-developing field of 3D imaging techniques, light-field (or \textit{plenoptic}) imaging is one of the most promising and used.
Light-field devices can measure simultaneously, without the need for moving parts, both the light distribution from the scene of interest and the propagation direction of light rays \cite{PI,8022901,Lam:15}. 
The availability of directional information makes single-shot volumetric reconstruction possible, whereas a standard imaging system would require multiple acquisitions on a series of independent axial planes to obtain a similar amount of information. 

One of the main reasons behind the success of light-field imaging is the structural simplicity of the devices, essentially consisting of standard cameras integrated with an array of microlenses between the main lens and the sensor. Such a structure ensures contained costs and high acquisition speed, which makes light-field imaging popular in the most diverse fields, ranging from photography to microscopy \cite{PI1,PI2,Broxton:13}, for cutting-edge application such as imaging of neuronal activity \cite{PI6} and wavefront sensing \cite{Ko:17}.
Nonetheless, traditional light-field devices such as those described above suffer from a limitation in the best resolution that can be achieved, due to the fact that a \textit{single} sensor captures a composite information, in which both the spatial distribution and the direction of light are encoded \cite{PI7,PI4}.

An alternative method to light-field imaging capable of addressing the resolution loss has recently emerged in the context of correlation imaging \cite{pittman1995optical,bennink2002two,valencia2005two,gatti2004ghost,scarcelli2006can,osullivan2010comparison,brida2011systematic}. In this approach, called correlation plenoptic imaging (CPI), the plenoptic information is spit over two separate sensors, both endowed with spatial resolution \cite{DAngelo2016,CPIent,Pepe2017}. While a simple intensity measurement on each detector does not contain volumetric information, the latter is encoded in the correlation between intensity fluctuations registered at each pair of pixels of the disjoint sensors. 
Besides recovering Rayleigh-limited resolution of focused images, CPI has further advantages deriving from an unparalleled extension of the longitudinal depth along which the sample can be correctly reconstructed starting from a single plenoptic image \cite{overview,Scagliola2020,DiLena2020}.
Since its introduction, great improvement to the performance of the device have been carried over, both in terms of its optical performance \cite{Scagliola2020,DiLena2020} and signal-to-noise ratio optimization \cite{scala,Massaro2022_SNR}, to obtain remarkable performance in microscopy applications \cite{massaro2023lightfield} and in acquisition speed \cite{massaro2023quantum}.

In this article, we present a novel CPI configuration, based on measuring correlations between position and momentum measurements: the first one being measured in points on a given object plane, close to the imaged sample, and the second one in points of the Fourier plane of the imaging lens \cite{goodman2005introduction}. Previous CPI setups were based on the fact that intensity correlations encode images of the same scene, as if it were illuminated by different point sources. Compared to microlens-based light-field techniques (even using correlation measurements in so-called ghost imaging, see Ref.~\cite{paniate2023light}), CPI potentially provides a much wider variety of independent viewpoints on a 3D sample. However, such an interesting property has practical drawbacks, such as the sensitivity to the axial distances of planes that are difficult to identify and the dependence of image magnification on the axial position (see Ref.~\cite{overview}). Here, we will show how using position-momentum correlation measurement, besides its fundamental interest, also helps to overcome even such technical problems of CPI. 

The article is organized as follows. In Section~\ref{sec:properties}, we illustrate the plenoptic capability of the proposed correlation measurement protocol, offering a
theoretical demonstration of its working principle. In Section~\ref{sec:performance}, we characterize the refocusing capability of the protocol by using Gaussian test objects and a numerical simulation applied to a planar target. The capability of the proposed CPI scheme to overcome the mentioned drawbacks of previous CPI protocols is discussed in Section~\ref{sec:discussion}, together with a summary of the main results.

\section{Plenoptic properties of intensity correlations}\label{sec:properties}

\begin{figure}
\begin{centering}
\includegraphics[width=0.8\textwidth]{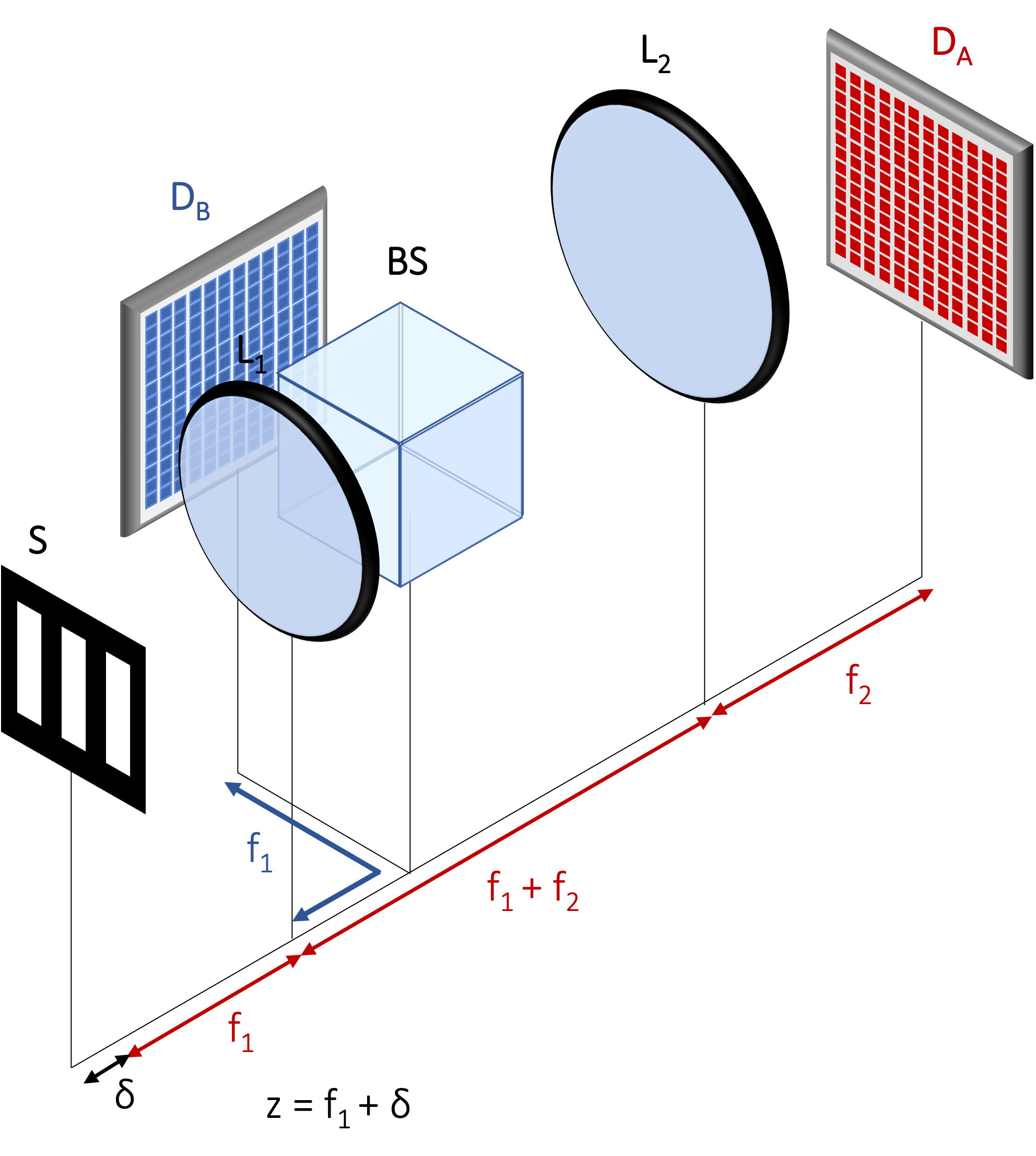}
\par\end{centering}
\centering{}\caption{A schematic representation of the setup: the object, represented as
a flat triple slit \textit{S} and treated as a chaotic light emitter,
is placed at a distance $z=f_{1}+\delta$ from the first lens $L_{1}$ of a two-lenses system, in a \emph{f-f configuration}, composed of
$L_{1}$ with focal length $f_{1}$ and \emph{$L_{2}$}
with focal length $f_{2}$; the two lenses are separated by a distance
$f_{1}+f_{2}$. The detector $D_{A}$ is in the second focal plane
of the second lens, which is, in the conjugate plane of the first focal plane of
$L_{1}$. In the space between the two lenses, a beam splitter \emph{BS}
deviates half of the intensity in the second arm of the setup toward the
detector $D_{B}$, which is placed in the Fourier plane of the first lens.\label{fig:setup}}
\end{figure}

A simple setup for measuring correlations between the near-field and the far-field of a sample is schematically
represented in Fig.~\ref{fig:setup}. Light emitted from an object
\emph{S} propagates through a two-lens system made of a first converging
lens\emph{ $L_{1}$ }with focal length $f_{1}$ and a second converging
lens \emph{$L_{2}$} with focal length $f_{2}$ and is collected by
a detector $D_{A}$ placed in the second focal plane of the second
lens. This detector, endowed with spatial resolution, collects the
image of the intensity distribution on the first focal plane of $L_{1}$,
reversed and magnified by a factor $M=f_{2}/f_{1}$. A second detector
$D_{B}$ is placed directly in the second focal plane of $L_{1}$ (i.e., in the  Fourier plane), giving access to far-field information; light is then deflected towards $D_{B}$ by means of a beam splitter
(BS) placed between the two lenses, so that part of the intensity
coming from the first lens in $D_{B}$ is moved in an orthogonal path.
Let us call $I_{A}(\boldsymbol{\rho_{A}})$ and $I_{B}(\boldsymbol{\rho_{B}})$
the intensity distributions collected by $D_{A}$ and $D_{B}$, respectively,
where $\boldsymbol{\rho_{A}}$ and $\boldsymbol{\rho_{B}}$ are two-dimensional
transverse spatial coordinates defined on the photo-sensitive planes
of the two detectors. It is worth noticing that the main lens Fourier plane, in cases when it is  physically inaccessible such as in most microscope objectives, can be imaged on a farther plane by using a relay imaging system.

By correlating, pixel by pixel, the signal at the two detectors, a \emph{position-momentum}
correlation is measured. If the object emits (or transmits
or reflects) thermal light \cite{mandel_wolf_1995}, this correlation contains plenoptic
information \cite{DAngelo2016}. In general, the relevant information
is contained in the second order correlation function $\Gamma^{(2)}$,
the four dimensional function representing the correlations between
the intensity fluctuations defined as 
\begin{equation}
\Gamma^{(2)}(\boldsymbol{\rho}_{\mathbf{A}},\boldsymbol{\rho_{B}})=\langle I_{A}(\boldsymbol{\rho_{A}})I_{B}(\boldsymbol{\rho_{B}})\rangle-\langle I_{A}(\boldsymbol{\rho_{A}})\rangle\langle I_{B}(\boldsymbol{\rho_{B}})\rangle\label{eq:gamma_general}
\end{equation}
where the $\langle\dots\rangle$ denotes the ensemble average over
the different statistical realizations of the intensity distributions. To better visualize the plenoptic capabilities contained
in $\Gamma^{(2)}$ and how these can lead to refocusing out-of-focus
objects, we consider a two-dimensional object with emission profile $\mathcal{A}(\boldsymbol{\rho_{s}})$, shining quasi-monochromatic
thermal light on the optical apparatus, placed at a distance $z$ from the first lens $L_{1}$.
By using the paraxial optics transfer functions to propagate the fields \cite{goodman2005introduction},
and assuming negligible coherence area \cite{saleh2013fundamentals} at the object plane
(i.e., the object is considered as a source of incoherent light), the second order
correlation function can be calculated and reads, up to irrelevant
constant factors:
\begin{multline}
\Gamma^{(2)}(\boldsymbol{\rho_{a}},\boldsymbol{\rho_{b}})\sim\Biggl|\int_{-\infty}^{+\infty}d^{2}\boldsymbol{\rho_{s}}|\mathcal{A}(\boldsymbol{\rho_{s}})|^{2}\int_{-\infty}^{+\infty}d^{2}\boldsymbol{\rho_{1}}\mathcal{P}_{L1}(\boldsymbol{\rho_{1}}) \\
\times \int_{-\infty}^{+\infty}d^{2}\boldsymbol{\rho'_{1}}\mathcal{P}_{L1}(\boldsymbol{\rho'_{1}})e^{i\phi(\boldsymbol{\rho_{s}},\boldsymbol{\rho_{1}},\boldsymbol{\rho'_{1}},\boldsymbol{\rho_{A}},\boldsymbol{\rho_{B}})}\Biggr|^{2}\label{eq:gamma}
\end{multline}
where we have considered the aperture of the lens $L_{1}$, as described
by the pupil function $\mathcal{P}_{L1}(\boldsymbol{\rho_{1}})$ (namely, a real function representing the geometrical shape of the lens), to be the relevant aperture
of the system, thus neglecting the one of the second lens, treated as infinite.
The term $\phi$ calculated in the coordinates of the sample, the
first lens and the detectors, is characteristic of the specific protocol
and reads 
\begin{equation}
\phi(\boldsymbol{\rho_{s}},\boldsymbol{\rho_{1}},\boldsymbol{\rho'_{1}},\boldsymbol{\rho_{A}},\boldsymbol{\rho_{B}})=k\Big[\Big(\frac{1}{z}-\frac{1}{f_{1}}\Big)\frac{\boldsymbol{\rho_{1}^{\textrm{2}}}}{2}-\boldsymbol{\rho_{1}}\cdot\Big(\frac{\boldsymbol{\rho_{s}}}{z}+\frac{\boldsymbol{\rho_{A}}}{f_{2}}\Big)-\frac{\boldsymbol{\rho'{}_{1}^{\textrm{2}}}}{2z}+\boldsymbol{\rho'_{1}}\cdot\Big(\frac{\boldsymbol{\rho_{s}}}{z}+\frac{\boldsymbol{\rho_{B}}}{f_{1}}\Big)\Big]\label{eq:phase}
\end{equation}
where $k=\frac{2\pi}{\lambda}$ is the wavenumber and $\lambda$ the
central wavelength of the quasi-monochromatic emission. The integral
in Eq.~\eqref{eq:gamma} can be solved in the geometrical optics
limit using a stationary phase approximation \cite{asymp}. Calling $\boldsymbol{\rho_{j}}=(x_{j},y_{j})$,
with $j=A,B$, we can further simplify the result writing the $\Gamma^{(2)}$
as a two-variable function factorized in a geometrical part $\Gamma_{geom}^{(2)}(x_{A},x_{B})$
involving the object-related information leading to refocusing and
a factor involving apertures $\Gamma_{pupils}^{(2)}(x_{A},x_{B})=|\mathcal{P}_{L1}(x_{A},x_{B})|^{4}$
\cite{Massaro2022}. The first term can be written as
\begin{equation}
\Gamma_{geom}^{(2)}(x_{A},x_{B})=\left\lvert\mathcal{A}\left(-\frac{x_{A}}{M}+\left (1-\frac{z}{f_{1}}\right )x_{B}\right)\right\rvert^{4}.\label{eq:gamma_geom_z}
\end{equation}
The geometrical meaning of the correspondence between an object point
$x_{s}$ and the two detector coordinates, as contained in this function,
becomes evident by inserting a real parameter $z=f_{1}+\delta$, with $\delta$ the defocusing parameter, defining the out-of-focus distance at
which the object is placed; with this substitution, Eq.~\eqref{eq:gamma_geom_z}
becomes
\[
\Gamma_{geom}^{(2)}(x_{A},x_{B})=\left\lvert\mathcal{A}\Big(-\frac{x_{A}}{M}-\delta\frac{x_{B}}{f_{1}}\Big)\right\rvert^{4}.
\]
In this expression, we can recognize the quantity $x_{B}/f_{1}$, given by the ratio between
the position of a point in the Fourier plane and the focal length
of the first lens $L_{1}$, as representing the angle of propagation from the object plane to the lens. By fixing this coordinate, all rays propagating at this angle from all the object points can be reconstructed. Concurrently, the term $-x_{A}/M$ represents a
point in the first focal plane of $L_{1}$. Fixing the coordinates
$x_{B}$ and $x_{A}$ uniquely selects, among the rays propagating
at the angle $x_{B}/f_{1}$, the one passing through $-x_{A}/M$. The
knowledge of the parameter $\delta$ completes the identification of
a single point $x_{s}$ on the object. Formally, this correspondence
represents a line ($\gamma_{s}(x_{s}):-\frac{x_{A}}{M}-\frac{\delta}{f_{1}}x_{B}=x_{s}$)
in the $(x_{A},x_{B})$ space, along which it is convenient to integrate
the $\Gamma^{(2)}(x_{A},x_{B})$ to reconstruct the refocused image \cite{s22176665, Massaro2022}.
In the geometrical approximation in which we are working, this general integration
can be written 
\begin{align}\label{eq:refoc}
\Sigma(x_{s}) & =\int_{\gamma_{s}(x_{s})}\Gamma^{(2)}(x_{A},x_{B})d\ell=\int_{\gamma_{s}(x_{s})}\Gamma_{geom}^{(2)}(x_{A},x_{B})\Gamma_{pupils}^{(2)}(x_{A},x_{B})d\ell \nonumber \\
& =\int_{\gamma_{s}(x_{s})}\left\lvert\mathcal{A}\Big(-\frac{x_{A}}{M}-\frac{\delta}{f_{1}}x_{B}\Big)\right\rvert^{4}\left\lvert\mathcal{P}_{L1}(x_{A},x_{B})\right\rvert^{4}d\ell\sim\left\lvert\mathcal{A}(x_{s})\right\rvert^{4},
\end{align}
considering that the artifacts to the object reconstruction created
by the presence of the $\Gamma_{pupils}^{(2)}(x_{A},x_{B})$ term
can easily be corrected \cite{Massaro2022}. 

It is worth noticing that the magnification of all the object sub-images, obtained for each $x_B$ and concurring to the formation of the refocused image, is independent of the object axial position and coincides with the native magnification $M$ of focused images. Such a feature, unique in the context of CPI, ensures an increased homogeneity of the image properties of axially extended samples.

\section{Imaging performance of CPI based on position-momentum measurements}\label{sec:performance}

The adoption of the geometrical approximation is convenient to explicitly retrieve both the correspondence
between detector and object coordinates encoded in $\Gamma^{(2)}$ and the lines along which to integrate to obtain refocusing. However, this approach  does not give evidence of the wave-optics effects of propagation and diffraction resulting in the definition of resolution and depth of field. To quantitatively show how CPI refocusing translates into an enhancement of the depth of field with respect to standard imaging, we consider a Gaussian object with intensity profile  $|\mathcal{A}(x_{s})|^{2}=\exp\big({-\frac{x_{s}^{2}}{2\sigma_{s}^{2}}}\big)$ and standard deviation $\sigma_{s}$ defined by its linear size,
and place it at a distance $\delta$ from the first focal plane of $L_{1}$;
we then calculate both its refocused image for the specific $\delta$ (as given by the proposed CPI protocol, but without the application of the stationary phase
approximations) and its standard image, as acquired directly
by the detector $D_{A}$. The spreading in transverse dimension
of both these images is finally evaluated for varying $\delta$ and $\sigma_{s}$. 

\begin{figure}
\begin{centering}
\includegraphics[scale=0.7]{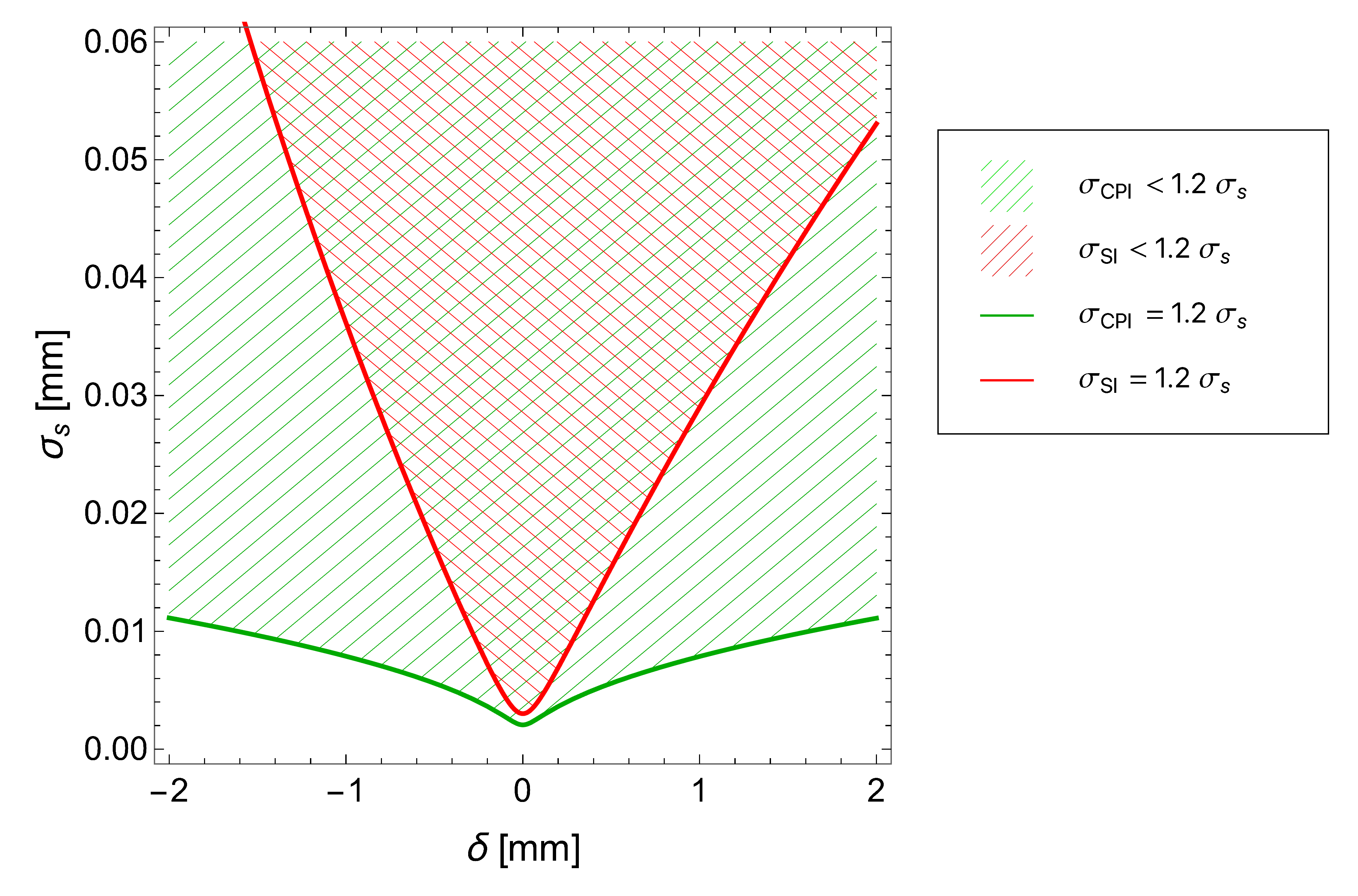}
\par\end{centering}
\caption{An analitycal resolution against depth of field comparison between
the CPI protocol discussed in the text and standard imaging, performed
calculating how should vary the size of a Gaussian intensity object
of width $\sigma_{s}$ and its distance from the focal plane $\delta$,
to be imaged with a width below a certain threshold. In particular,
the region filled with tilted green lines indicates the values of
$\sigma_{s}$ and the positions $\delta$ producing refocused images
with a linear size $\sigma_{CPI}$ that is less than $20\%$ of the
original object size while the region filled with red tilted lines
the values of $\sigma_{s}$ and $\delta$ producing standard images
of size $\sigma_{SI}$ less than the $20\%$ of the object width.
The green and red solid lines represent the boundaries of these regions
on which the image size is exactly $1.2$ times $\sigma_{s}$, for
CPI and standard imaging, respectively. The presented curves are calculated
for $M=1$, $f_{1}=f_{2}=30\;\mathrm{mm}$, $\sigma_{1}=0.635\;\mathrm{mm}$, $\lambda=532\;\mathrm{nm}$
and sample characterized by full transversely incoherent emission.\label{fig:resolution}}
\end{figure}

To obtain the refocused image analitically, we assume a Gaussian approximation
for the pupil function ($|\mathcal{P}_{L1}(x_{1})|^{2}=\exp\big({-\frac{x_{1}^{2}}{2\sigma_{1}^{2}}}\big)$)
and calculate $\Gamma^{(2)}$ by means of Eq. \eqref{eq:gamma}; we then 
apply the refocusing integration of Eq. \eqref{eq:refoc} and obtain 
again a Gaussian function with standard deviation $\sigma_{CPI}(\sigma_{s},\delta)$
dependent on both the size and the position of the object. Using the paraxial
propagators of the electric field, we calculate the standard image
acquired by the detector $D_{A}$ and we call $\sigma_{SI}(\sigma_{s},\delta)$
its transverse dimension also dependent on $\sigma_{s}$ and $\delta$.
To perform a comparison, we conventionally consider, for each defocusing $\delta$
and width $\sigma_{s}$, the regions in which $\sigma_{CPI}(\sigma_{s},\delta)$
and $\sigma_{SI}(\sigma_{s},\delta)$ are smaller than $1.2\;\sigma_{s}$ (\textit{i.e.}\ the image spread is smaller than $20\%$ of the
original object size). The results are presented in Fig.~\ref{fig:resolution},
where the vertical and horizontal axes report, respectively, the transverse dimensions of the object $\sigma_{s}$
and the defocusing parameter $\delta$ at which the object is placed: the green and red solid contour lines represent
the threshold on which the transverse dimension of the image is broadened by $20\%$ due to loss of resolution (namely, when the image size becomes $1.2$ times the dimension of the object), for CPI and standard imaging
respectively; the green and red tilted lines fill the regions
in which the dimension of the images is below this threshold.

This plot can be interpreted as an indication of the different behavior of resolution as the object is moved from the focal plane, in the two imaging technques. Similar trends can be observed in the visibility-based resolution limits described in Refs.~\cite{Pepe2017,Scagliola2020}. Notice, however, that curves in Fig.~\ref{fig:resolution} do not directly carry information about the visibility of the image of two separated object points: $\sigma_{s}$
can be viewed as the detail size on the object side and the colored
regions indicate, for each distance from the focal plane, the detail size that can be imaged by the two protocols with a spreading
below a certain threshold, which is set to 20\% as an arbitrary reference. The trends in Fig.~\ref{fig:resolution} provide a qualitative indication of the loss of resolution with distance from the focal plane.

The comparison between the two protocols, in particular, shows how, outside of the natural depth of field of the system, while for standard imaging the degradation
of resolution scales approximately linearly with the distance from
the focal plane, exploiting correlations between position and momentum
allows this trend to become a square root of the same distance, leading
to a slower loss of details, typical of CPI protocols \cite{Pepe2017,Scagliola2020,DiLena2020}.

It is finally worth pointing out the difference in behavior of
the two protocols in the focal plane of $L_{1}$ for $\delta=0$:
the lower spreading (hence, better resolution) associated to CPI is due to
the second order correlation measurement in the process
of image formation and should not be interpreted as a characteristic
of the optical system. The correlation-based imaging process, as can
be clearly understood by the third integral of Eq.~\eqref{eq:refoc},
involves the integration over the second power of the intensity of
the object whose size, for a Gaussian object, is $1/\sqrt{2}$ the
size of the object itself. This means that, for Gaussian objects, the
variance of the produced image is the sum of the squares of a diffractive
term, which depends on the numerical aperture of the system and on the
wavelength $\lambda$, and the original object size divided by a factor
$\sqrt{2}$; for standard imaging, the diffractive term is the
same, but the second term in the sum is the object size $\sigma_{s}$ itself. 

To show an example of the refocusing properties of the position-momentum
based CPI protocol, we simulate
the measurement of a correlation function in the experimental situation
discussed above and apply the refocusing procedure reported in
Eq.~\eqref{eq:refoc} to obtain refocusing. 
In particular, we consider a two-dimensional object placed at a distance
$\delta=+1\;\mathrm{mm}$ away from the first focal plane of the first lens $L_{1}$ (plane at focus). The object is a mask composed by three perfectly transmissive rectangular
slits, having center-to-center distance of $50\;\mu\mathrm{m}$.\\
The incoherent emission from the target is simulated by computing a series
of speckle patterns \cite{speckle} following pseudo-random thermal statistics
and propagating them
along the two arms of the setup to reach the two detectors (with no modifications to the statistics of the incoming light due to the presence of the beam splitter), and then obtaining the two intensity distributions
$I_{A}(\boldsymbol{\rho_{A}})$ and $I_{B}(\boldsymbol{\rho_{B}})$ by collecting \emph{N} statistically independent frames.
By using Eq.~\eqref{eq:gamma_general} and an ergodic hypothesis, we are now able to calculate the correlation function;
the simulated second order correlation function shall be indicated as  $\Gamma_{sim}^{(2)}$.

\begin{figure}
\begin{centering}
\includegraphics[width=0.8\textwidth]{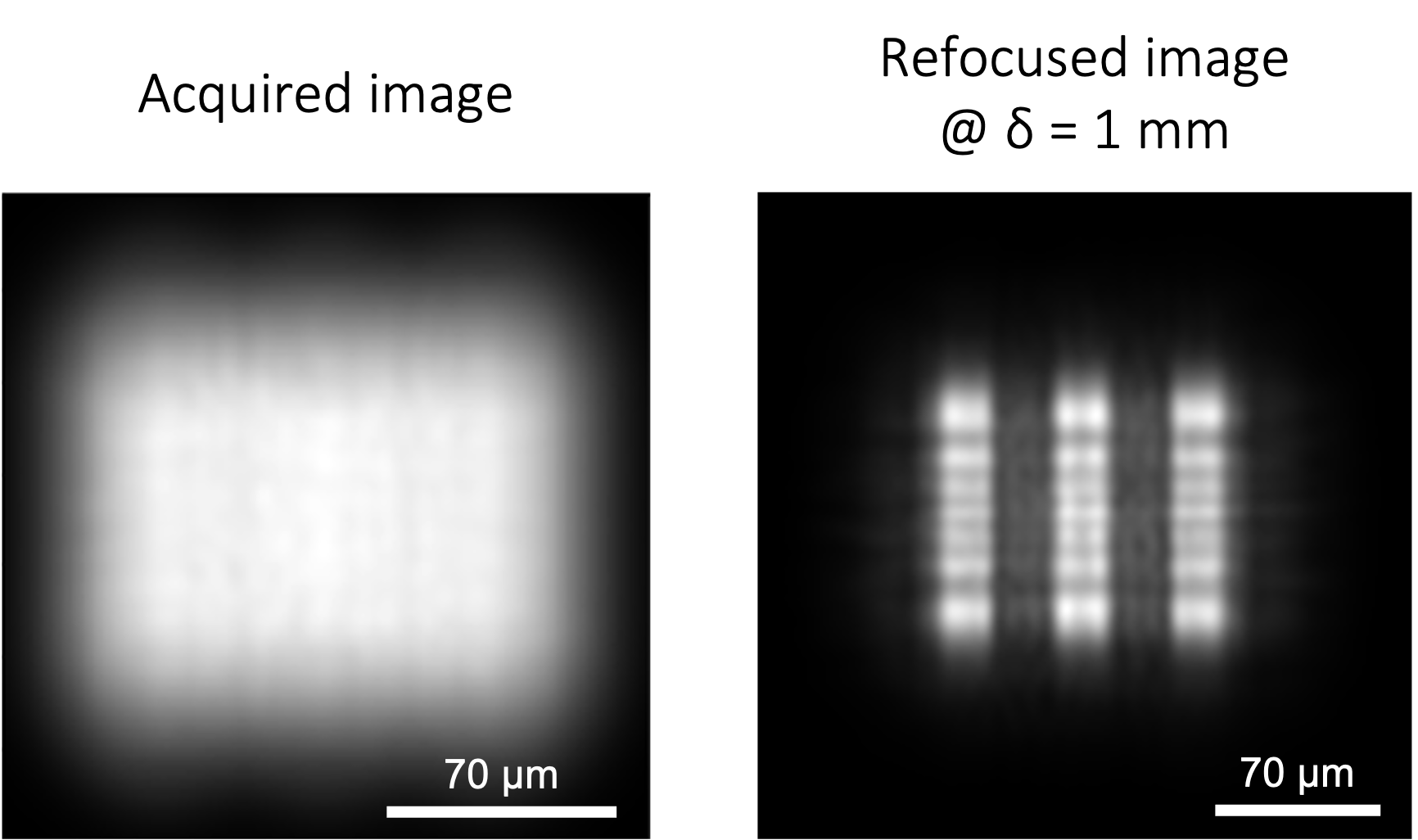}
\par\end{centering}
\caption{Left: simulation of the intensity acquired by
the detector $D_{A}$ over the collected $N$ frames, representing
the completely blurred image of the mask placed at $\delta=1$ mm
from the \emph{$L_{1}$} lens first focal plane, outside the depth
of field of the two-lenses system. Right: refocused image of the same
mask placed in the same position, reconstructed by exploiting the plenoptic
information contained in $\Gamma_{sim}^{(2)}$, the correlation
function calculated from the simulated frames collected by the two
detectors $D_{A}$ and $D_{B}$, as retrieved after the application of
the first integral of Eq.~\eqref{eq:refoc}. The scale bar indicates
approximately $70\;\mu$m. Notice that the parameters of the main lens used in the simulation are matched with those of the Gaussian lens used for Fig.~\ref{fig:resolution}, by requiring that the two lenses yield the same visibility when imaging two separated point sources on the focus plane.}\label{fig:refocused}
\end{figure}

In the left panel of Fig.~\ref{fig:refocused}, we report the calculated standard
image as acquired by the detector $D_{A}$: the triple slit
appears completely blurred since the mask is placed far beyond the natural
depth of field of the two-lens system. The image is obtained by simply
averaging all the $I_{A}(\boldsymbol{\rho_{A}})$ distributions
acquired over the $N$ frames. In the right panel, we report the perfectly resolved 
refocusing of the object performed by taking advantage
of the plenoptic content of the correlation function, as obtained by
applying directly the first integral of Eq.~\eqref{eq:refoc} to $\Gamma_{sim}^{(2)}$. In the numerical simulation, the magnification
of the two-lens system is set to $M=1$, the lens $L_{1}$ acts
as a thin lens with numerical aperture $\text{NA}$ = 0.05 and focal length
$f_{1}=30\,$mm and the chosen number of frames is $N=15000$. 

\section{Summary and discussion}\label{sec:discussion}

We have presented a novel protocol to perform correlation plenoptic
imaging based on the measurement of correlations between intensity
fluctuations occurring in the object plane and in the Fourier plane
of a two-lens system, \textit{i.e.}\ exploiting position-momentum
correlations. Starting with a geometrical approximation, we have demonstrated
the plenoptic capability of the protocol and derived the so-called
\textit{refocusing algorithm}, namely, the coordinate transformation to be
applied to the measured correlation function $\Gamma^{(2)}$; such transformation  is just the path
over which performing the linear integration of $\Gamma^{(2)}$ to recover
the image of the out-of-focus object. The trend of the resolution as a function of the depth of field of the proposed CPI protocol has highlighted the square-root
scaling responsible for the wide depth of field enhancement of CPI (compared to
the linear scaling of standard imaging). The presented numerical simulation has also shown the result obtained by refocusing the image of $50\,\mu\mathrm{m}$ slit, placed $1\,\mathrm{mm}$ out of focus, outside
the region of resolvable details for standard imaging. This completes
the demonstration of the functionality of the proposed correlation plenoptic
imaging scheme, confirming the flexibility of CPI protocols in the
choice of planes over which measuring correlations. 

An outstanding feature of the proposed protocol is the fact that the sub-images obtained by selecting different illumination direction are characterized by the same magnification, regardless of the axial distance from the main lens. Such a feature, bearing an interesting parallelism with telecentric imaging systems, ensures a much larger homogeneity of the refocused images along the axial direction, which was not incorporated in the previous CPI schemes, where \textit{points} of illumination, instead of \textit{directions}, can be naturally selected from correlations of intensity fluctuations.

Moreover, it is worth noticing that the optical performance of the proposed implementation is similar to the one characterizing the microscopy-oriented CPI setup discussed in Ref.~\cite{massaro2023lightfield}), where correlations are measured between the near-field of the object, imaged through a conventional microscope, and the light intensity on the first lens surface.
Although theoretically well-defined in the thin-lens approximation, the latter configuration is not as simple to reproduce experimentally when thick lenses are involved. In these cases, the imaged plane is either chosen between the lens principal planes, or any other plane within the optical component. However, these planes are typically not easily accessible experimentally, do not always show distinctive optical features, and can only be determined with a certain degree of uncertainty. For these reasons, the refocusing procedure needs accurate calibration in order to determine the appropriate coefficients.
Imaging the Fourier plane as we propose here, instead, removes all the uncertainties in this regard, since the second focal plane is always well-defined even for real optical components and is easily recognized experimentally due to its distinctive optical features.

\section*{Author contributions}
\noindent Conceptualization: all authors; methodology: D.G., G.M., M.D., F.V.P.; software: G.M.; formal analysis: D.G., G.M.; validation: B.S., M.D., F.V.P.; writing--original draft preparation: D.G., G.M.; writing--review and editing: all authors; visualization: D.G., G.M.; supervision: F.V.P., M.D.; project administration: M.D.; funding acquisition: M.D., B.S. All authors have read and agreed to the submitted version of the manuscript.

\section*{Acknowledgments}
\noindent The research is supported by project Qu3D, funded by the Italian Istituto Nazionale di Fisica Nucleare, the Swiss National Science Foundation (grant 20QT21 187716 ``Quantum 3D Imaging at high speed and high resolution''), the Greek General Secretariat for Research and Technology, the Czech Ministry of Education, Youth and Sports, under the QuantERA programme, which has received funding from the European Union's Horizon 2020 research and innovation programme. D.G., G.M., M.D. and F.V.P. are supported by Istituto Nazionale di Fisica Nucleare (INFN) through project QUISS.  M.D. is supported by European Union-NextGenerationEU PE0000023 - ``National Quantum Science and Technology Institute''. F.V.P. is supported by European Union-NextGenerationEU CN00000013 - ``National Centre for HPC, Big Data and Quantum Computing''.


\begin{thebibliography}{10}
\expandafter\ifx\csname url\endcsname\relax
  \def\url#1{\texttt{#1}}\fi
\expandafter\ifx\csname urlprefix\endcsname\relax\def\urlprefix{URL }\fi
\expandafter\ifx\csname href\endcsname\relax
  \def\href#1#2{#2} \def\path#1{#1}\fi

\bibitem{PI}
E.~H. Adelson, J.~Y.~A. Wang, Single lens stereo with a plenoptic camera, {IEEE
  Trans. Pattern Anal. Mach. Intell.} 14 (1992) 99--106.

\bibitem{8022901}
G.~Wu, B.~Masia, A.~Jarabo, Y.~Zhang, L.~Wang, Q.~Dai, T.~Chai, Y.~Liu, {Light
  Field Image Processing: An Overview}, {IEEE J. Sel. Top. Signal Process.} 11
  (2017) 926--954.

\bibitem{Lam:15}
E.~Y. Lam, Computational photography with plenoptic camera and light field
  capture: tutorial, {J. Opt. Soc. Am. A} 32 (2015) 2021--2032.

\bibitem{PI1}
R.~NG, M.~{Levoy}, M.~{Br{\'e}dif}, G.~{Duval}, M.~{Horowitz}, Pat{ Hanrahan},
  Light field photography with a hand-held plenoptic camera, Stanford
  University Computer Science Tech Report CSTR 2005-02 (April 2005).

\bibitem{PI2}
R.~Ng, {Fourier Slice Photography}, ACM Trans. Graph. 24 (2005) 735--744.

\bibitem{Broxton:13}
M.~Broxton, L.~Grosenick, S.~Yang, N.~Cohen, A.~Andalman, K.~Deisseroth,
  M.~Levoy, Wave optics theory and 3-d deconvolution for the light field
  microscope, Opt. Express 21 (2013) 25418--25439.

\bibitem{PI6}
R.~Prevedel, Y.-G. Yoon, M.~Hoffmann, N.~Pak, G.~Wetzstein, S.~Kato,
  T.~Schr{\"o}del, R.~Raskar, M.~Zimmer, E.~S. Boyden, A.~Vaziri, Simultaneous
  whole-animal 3d imaging of neuronal activity using light-field microscopy,
  Nat. Methods 11 (2014) 727--730.

\bibitem{Ko:17}
J.~Ko, C.~C. Davis, Comparison of the plenoptic sensor and the shack-hartmann
  sensor, {Appl. Opt.} 56 (2017) 3689--3698.

\bibitem{PI7}
T.~Georgeiv, K.~C. Zheng, B.~Curless, D.~Salesin, S.~K. Nayar, C.~Intwala,
  Spatio-angular resolution tradeoffs in integral photography, in: EGSR '06:
  Proceedings of the 17th Eurographics conference on Rendering Techniques,
  2006, pp. 263--272.

\bibitem{PI4}
B.~Goldluecke, O.~Klehm, S.~Wanner, E.~Eisemann, Plenoptic Cameras, CRC Press,
  2015, Ch. Digital Representations of the Real World: How to Capture, Model,
  and Render Visual Reality, pp. 65--77.

\bibitem{pittman1995optical}
T.~B. Pittman, Y.-H. Shih, D.~V. Strekalov, A.~V. Sergienko, Optical imaging by
  means of two-photon quantum entanglement, Phys. Rev. A 52 (1995) R3429.

\bibitem{bennink2002two}
R.~S. Bennink, S.~J. Bentley, R.~W. Boyd, {``Two-photon'' coincidence imaging
  with a classical source}, Phys. Rev. Lett. 89 (2002) 113601.

\bibitem{valencia2005two}
A.~Valencia, G.~Scarcelli, M.~D'Angelo, Y.~Shih, Two-photon imaging with
  thermal light, Phys. Rev. Lett. 94 (2005) 063601.

\bibitem{gatti2004ghost}
A.~Gatti, E.~Brambilla, M.~Bache, L.~A. Lugiato, Ghost imaging with thermal
  light: comparing entanglement and classical correlation, Phys. Rev. Lett. 93
  (2004) 093602.

\bibitem{scarcelli2006can}
G.~Scarcelli, V.~Berardi, Y.~Shih, Can two-photon correlation of chaotic light
  be considered as correlation of intensity fluctuations?, Phys. Rev. Lett. 96
  (2006) 063602.

\bibitem{osullivan2010comparison}
M.~N. O'Sullivan, K.~W.~C. Chan, R.~W. Boyd, Comparison of the signal-to-noise
  characteristics of quantum versus thermal ghost imaging, Phys. Rev. A 82
  (2010) 053803.

\bibitem{brida2011systematic}
G.~Brida, M.~Chekhova, G.~Fornaro, M.~Genovese, E.~Lopaeva, I.~R. Berchera,
  Systematic analysis of signal-to-noise ratio in bipartite ghost imaging with
  classical and quantum light, Phys. Rev. A 83 (2011) 063807.

\bibitem{DAngelo2016}
M.~D'Angelo, F.~V. Pepe, A.~Garuccio, G.~Scarcelli, Correlation plenoptic
  imaging, Phys. Rev. Lett. 116 (2016) 223602.

\bibitem{CPIent}
F.~Pepe, F.~Di~Lena, A.~Garuccio, G.~Scarcelli, M.~D'Angelo, {Correlation
  Plenoptic Imaging With Entangled Photons}, Technologies 4 (2016).

\bibitem{Pepe2017}
F.~V. Pepe, F.~Di~Lena, A.~Mazzilli, E.~Edrei, A.~Garuccio, G.~Scarcelli,
  M.~D'Angelo, Diffraction-limited plenoptic imaging with correlated light,
  {Phys. Rev. Lett.} 119 (2017) 243602.

\bibitem{overview}
F.~Di~Lena, F.~Pepe, A.~Garuccio, M.~D'Angelo, {Correlation Plenoptic Imaging:
  An Overview}, Appl. Sci. 8 (2018) 1958.

\bibitem{Scagliola2020}
A.~Scagliola, F.~Di~Lena, A.~Garuccio, M.~D'Angelo, F.~V. Pepe, Correlation
  plenoptic imaging for microscopy applications, Phys. Lett. A 384 (2020)
  126472.

\bibitem{DiLena2020}
F.~D. Lena, G.~Massaro, A.~Lupo, A.~Garuccio, F.~V. Pepe, M.~D'Angelo,
  Correlation plenoptic imaging between arbitrary planes, {Opt. Express} 28
  (2020) 35857--35868.

\bibitem{scala}
G.~Scala, M.~D'Angelo, A.~Garuccio, S.~Pascazio, F.~Pepe, Signal-to-noise
  properties of correlation plenoptic imaging with chaotic light, Phys. Rev. A
  99 (2019) 053808.

\bibitem{Massaro2022_SNR}
G.~Massaro, G.~Scala, M.~D'Angelo, F.~V. Pepe, Comparative analysis of
  signal-to-noise ratio in correlation plenoptic imaging architectures, Eur.
  Phys. J. Plus 137 (2022) 1123.

\bibitem{massaro2023lightfield}
G.~Massaro, D.~Giannella, A.~Scagliola, F.~Di~Lena, G.~Scarcelli, A.~Garuccio,
  F.~V. Pepe, M.~D'Angelo, Light-field microscopy with correlated beams for
  high-resolution volumetric imaging, Sci. Rep. 12 (2022) 16823.

\bibitem{massaro2023quantum}
G.~Massaro, P.~Mos, S.~Vasiukov, F.~D. Lena, F.~Scattarella, F.~V. Pepe,
  A.~Ulku, D.~Giannella, E.~Charbon, C.~Bruschini, M.~D'Angelo,
  Correlated-photon imaging at 10 volumetric images per second, Sci. Rep. 13
  (2023) 12813.

\bibitem{goodman2005introduction}
J.~W. Goodman, Introduction to Fourier optics, 3rd ed., Macmillan Learning,
  London, 2005.

\bibitem{paniate2023light}
A.~Paniate, G.~Massaro, A.~Avella, A.~Meda, F.~V. Pepe, M.~Genovese,
  M.~D'Angelo, I.~Ruo~Berchera, {Light Field Ghost Imaging}, arXiv preprint
  arXiv:2309.14701 (2023).

\bibitem{mandel_wolf_1995}
L.~Mandel, E.~Wolf, Optical Coherence and Quantum Optics, Cambridge University
  Press, Cambridge, 1995.

\bibitem{saleh2013fundamentals}
B.~Saleh, M.~Teich, Fundamentals of Photonics, John Wiley \& Sons, Hoboken, NJ,
  2013.

\bibitem{asymp}
R.~Wong, Asymptotic Approximations of Integrals, Society for Industrial and
  Applied Mathematics, 2001.

\bibitem{Massaro2022}
G.~Massaro, F.~Di~Lena, M.~D'Angelo, F.~V. Pepe, Effect of finite-sized optical
  components and pixels on light-field imaging through correlated light,
  Sensors 22 (2022) 2778.

\bibitem{s22176665}
G.~Massaro, F.~V. Pepe, M.~D'Angelo, Refocusing algorithm for correlation
  plenoptic imaging, Sensors 22 (2022) 6665.

\bibitem{speckle}
J.~W. Goodman, Speckle Phenomena in Optics: Theory and Applications, 2nd ed.,
  SPIE Press, Bellingham, WA, 2007.

\end{thebibliography}
\end{document}